\documentclass[12pt,preprint]{aastex}

\begin{document}
\newcommand{\msun}{M_{\odot}}
\newcommand{\kms}{\, {\rm km\, s}^{-1}}
\newcommand{\cm}{\, {\rm cm}}
\newcommand{\gm}{\, {\rm g}}
\newcommand{\erg}{\, {\rm erg}}
\newcommand{\kpc}{\, {\rm kpc}}
\newcommand{\mpc}{\, {\rm Mpc}}
\newcommand{\seg}{\, {\rm s}}
\newcommand{\kev}{\, {\rm keV}}
\newcommand{\hz}{\, {\rm Hz}}
\newcommand{\sr}{\, {\rm sr}}
\newcommand{\etal}{et al.\ }
\newcommand{\yr}{\, {\rm yr}}
\newcommand{\eq}{eq.\ }
\newcommand{\lya}{Ly$\alpha$\ }
\newcommand{\hi}{\mbox{H\,{\scriptsize I}\ }}
\newcommand{\hii}{\mbox{H\,{\scriptsize II}\ }}
\newcommand{\hei}{\mbox{He\,{\scriptsize I}\ }}
\newcommand{\heii}{\mbox{He\,{\scriptsize II}\ }}
\newcommand{\cii}{\mbox{C\,{\scriptsize II}\ }}
\newcommand{\ciis}{\mbox{C\,{\scriptsize II}${}^{\ast}$ }}
\newcommand{\nhi}{N_{HI}}
\def\arcsec{''\hskip-3pt .}

\title{
The Invisible Tension of the Universe from Astrophysical Black Holes:
A Solution to the Coincidence Problem of the Accelerated Expansion}
\author{Jordi Miralda-Escud\'e${}^{1,2}$}
\affil{${}^1$ Instituci\'o Catalana de Recerca i Estudis
 Avan\c cats, Barcelona, Catalonia.}
\affil{${}^2$ Institut de Ci\`encies del Cosmos, Universitat de Barcelona
 / IEEC, Barcelona, Catalonia.}

\begin{abstract}
  Astronomical observations have shown that the expansion of the
universe is at present accelerating, consistently with a constant
negative pressure or tension. This is a major puzzle because we do not
understand why this tension is so small compared to the Planck density;
why, being so small, it is not exactly zero; and why it has precisely
the required value to make the expansion start accelerating just at the
epoch when we are observing the universe. The recently proposed
conjecture by Afshordi that black holes create a gravitational aether
owing to quantum gravity effects, which may be identified with this
invisible tension, can solve this coincidence problem. The fact that the
expansion of the universe is starting to accelerate at the epoch when
we observe it is a necessity that is implied by our origin in a planet
orbiting a star that formed when the age of the universe was of the same
order as the lifetime of the star. This argument is unrelated to any
anthropic reasoning.
\end{abstract}

\section{Introduction}

  Observations of the Cosmic Microwave Background and other astronomical
distance determinations have convincingly demonstrated that the
expansion of the universe started accelerating at a recent epoch (see
Komatsu \etal 2010 and references therein). All of the available
observations are consistent with a cosmological constant, or the effect
of a constant negative pressure with value (we use Planck units with
$c=\hbar=G=1$ throughout this paper)
\begin{equation}
 p_{\Lambda} = - {3H_0^2\over 8\pi} \Omega_{\Lambda 0} =
 (1.3 \pm 0.1) \times 10^{-123} ~. 
\end{equation}
Here, $H_0$ is the value of the Hubble constant at present and
$\Omega_{\Lambda 0}$ is the ratio of the negative pressure to the
critical density of the universe, and we have used the values obtained
by Komatsu et al.\ As far as one can tell, this value of the invisible
tension of the universe appears to be a fundamental, dimensionless
constant of nature, unrelated to any other known physical law. Its
extremely small value, and the fact that this value is precisely the
required one to cause the acceleration of the expansion of the universe
to start at the epoch when we are observing it, is the major puzzle
that is referred to as the coincidence problem.

  There is only one other fundamental (and independent), dimensionless
physical constant of nature that is different from unity by many orders
of magnitude: this is related to the extreme weakness of the
gravitational interaction compared to all other fundamental interactions
(the ratio of neutrino to baryon masses may be considered another small
number, but we will ignore neutrino physics here). For example, the
ratio of the attractive gravitational and electric forces between a
proton and an electron (of masses $m_p$ and $m_e$) is
\begin{equation}
 {m_p m_e \over \alpha} = 4.4 \times 10^{-40} ~,
\end{equation}
where $\alpha$ is the fine structure constant of the electromagnetic
interaction.

  There are only two possibilities: either there is some relation
between these two very small constants of nature arising from some yet
unknown physical law, or there is not. If there is no relation, then we
have to reach the conclusion that our universe is characterized by two
numbers that are extremely small for two different reasons. The hope for
a simple description of the universe may lead one to suspect that there
is a relation, and if so, that an explanation for the coincidence
problem might be found in this relation.

  It is pointed out in this paper that this is precisely the implication
of the recent conjecture by Afshordi (2010), that the invisible tension
of the universe may be a result of quantum gravity effects from the
entropy of astrophysical black holes, which I outline briefly in \S 2.
The reason why this conjecture predicts that the expansion of the
universe starts accelerating when the age of the universe is of the
same order as the lifetime of a star is then explained in \S 3. This
argument was previously discussed in Miralda-Escud\'e (2007), where it
was presented as an April fool's day joke instead of the standard
scientific format, attributing the reason for a relation between the two
very small fundamental physical constants of nature to made-up nonsense,
as part of the joke.

\section{Gravitational aether and astrophysical black holes}

  Afshordi (2010; see also references therein) postulates a model of
emergent gravity, in which Lorentz symmetry is an emergent phenomenon at
low energies rather than a fundamental symmetry of nature. This
introduces a gravitational aether which, in the presence of quantum
corrections from a black hole horizon, may acquire a finite pressure. A
generic first-order finite temperature quantum correction for the
relation of the black hole entropy and mass to the temperature leads one
to conclude, via the first law of thermodynamics, that the quantum
correction implies a gravitational aether pressure
\begin{equation}
  p=-C_a \pi T_{bh}^3 ~,
\end{equation}
where $C_a$ is a dimensionless constant that would depend on the
quantum gravity theory, $T_{bh}$ is the black hole horizon temperature,
and the black hole mass is $m_{bh} = (8\pi T_{bh})^{-1}$.

  In the presence of many black holes in the universe, it was argued by
Prescod-Weinstein, Afshordi \& Balogh (2009) that the pressure of the
gravitational aether should settle to an approximate hydrostatic
equilibrium over most of the volume in between the black holes, to a
value that corresponds to a mean black hole mass, $\bar m_{bh}$, equal
to the mass-weighted geometric average of all the black holes in the
universe. Because most of the mass density in black holes is
contributed by the remnants of core collapse of massive stars, this
mean black hole mass should be not much larger than that of the typical
black hole formed from a single star, which is a few times the
Chandrasekhar mass.

\section{Solution to the coincidence problem}

  The Chandrasekhar mass, $m_{Ch}$, for a star made of $N_{Ch} =
m_{Ch}/m_p$ baryons, beyond which a degenerate star must collapse to a
black hole, is obtained from the condition that the degenerate energy of
the particles at the point when they become relativistic balances the
gravitational energy of the star. The maximum degeneracy energy of the
particles in a star of radius $R$ is $\sim N_{Ch}^{1/3}/R$, while the
gravitational energy is $\sim m_{Ch}^2/R$. Equating the two results in
$m_{Ch}\simeq m_p^{-2}$. Therefore, we can write $\bar m_{bh} =
C_{bh}/m_p^2$, where $C_{bh}$ is a constant that is not much larger than
one, since the typical black hole made by the collapse of stars has a
mass that is not much more than a few times the Chandrasekhar mass.

  The pressure of the gravitational aether is then
\begin{equation}
  p=- {C_a \pi m_p^6 \over (8\pi C_{bh})^3 } ~.
\end{equation}
This is the relation we have in this model between the two very small
constants of the universe, provided that $C_a$ and $C_{bh}$ are not
themselves extremely different from unity. The age of the universe when
this negative pressure becomes dynamically dominant for the expansion is
\begin{equation}
  t_e= \left( {3 \over 8 \pi | p | } \right)^{1/2} =
 { 8 \sqrt{3\pi} C_{bh}^{3/2} \over \sqrt{C_a} m_p^3 } ~.
\end{equation}
We are now interested in comparing this time to the lifetime of a star.
The lifetime can be expressed in terms of the efficiency of generating
energy from nuclear reactions, $\epsilon$, and the fraction of the
Eddington luminosity at which the star radiates energy, $\ell$. The
Eddington luminosity of a star of mass $M$ is given by
\begin{equation}
 L_{Edd} = {3\mu_e m_e^2 \over 2 \alpha^2} \, M ~,
\end{equation}
where $\mu_e$ is the mean mass per electron ($ \simeq 1.2 m_p$ for the
fully ionized primordial mixture of hydrogen and helium). The stellar
lifetime is
\begin{equation}
 t_s = {M \epsilon \over \ell L_{Edd}} = 
{2\alpha^2 \epsilon \over 3 \ell \mu_e m_e^2} ~.
\end{equation}
Hence, the ratio of the age of the universe to the stellar lifetime when
the cosmic acceleration starts is
\begin{equation}
 { t \over t_s } = {12 \sqrt{3\pi C_{bh}^3}\, \ell \mu_e m_e^2 \over 
 4 \alpha^2 \epsilon \sqrt{C_a} m_p^3 } ~.
\label{trat}
\end{equation}
This solves, at least in part, the coincidence problem: the ratio of the
age of the universe to the stellar lifetime does not depend on the
extremely small values of the particle masses (reflecting the weakness
of gravity when we use Planck units). It is therefore not so surprising
that the two times turn out to be comparable, even though they still
depend on quantities that are far from unity: the ratio of the electron
to proton mass, the fine-structure constant, the efficiency of nuclear
reactions, and the fraction of the Eddington luminosity at which a
certain star radiates. The fraction $\ell$ is close to unity for massive
stars, it is $\ell = 10^{-4.6}$ for the Sun, and drops to $\ell \sim
10^{-7}$ for the lowest mass stars that are still able to ignite nuclear
reactions.

\section{Discussion}

  The conjecture proposed by Afshordi (2010) that the invisible tension
in the universe arises from a gravitational aether that acquires a
negative pressure from the existence of astrophysical black holes
implies that this invisible tension scales as $m_p^{-6}$. This provides
an automatic explanation for why the acceleration of the expansion of
the universe is starting close to the epoch when we observe it. It is
still surprising that the combination of constants appearing in equation
(\ref{trat}) is roughly close to unity for a star like the Sun, but this
coincidence now seems much less unlikely than in the case when the two
very small fundamental constants of nature are unrelated.

  It should be noted that this explanation for the coincidence problem
is unrelated to the anthropic principle. The prediction that the
expansion should start accelerating when the age of the universe is of
the same order as the lifetime of a star is a purely physical one and
bears no relation to our presence in the universe. The fact that the
lifetime of the Sun and the present age of the universe are comparable
is known to be true, and it has not been considered a surprising
coincidence: this may be related to a ``weak''´ and obvious form of t
anthropic principle that says that we must appear in the universe at the
epoch when most of the stars adequate for harboring planets with life
are in their main-sequence phase.

\acknowledgements

  I acknowledge support by the Spanish grant AYA2009-09745.

\newpage
\centerline{ \bf REFERENCES }
\vskip 0.5cm
\noindent
Afshordi, N. 2010, arXiv:1003.4811. \par
\noindent
Komatsu, E., \etal 2010, submitted to ApJS, arXiv:1001.4538. \par
\noindent
Miralda-Escud\'e, J. 2007, astro-ph/0703774. \par
\noindent
Prescod-Weinstein, C., Afshordi, N., \& Balogh, M. L. 2009, Phys. Rev. D80, 043513 (arXiv:0905.3551). \par


\end{document}